\documentclass[aps,prd,superscriptaddress,amsmath,amssymb,twocolumn,fleqn,nofootinbib]{revtex4-1} 

\usepackage{xcolor}
\usepackage{graphicx}
\usepackage{colordvi}
\usepackage{mathtools}
\usepackage{empheq}
\usepackage{bm}%

\usepackage{floatrow}
\usepackage{braket}
\usepackage{soul}

\usepackage{amssymb}
\usepackage{amsmath}

\usepackage{hyperref}
\def\be{\begin{equation}}
\def\ee{\end{equation}}
\def\ba{\begin{eqnarray}}
\def\ea{\end{eqnarray}}

\newcommand{\dd}{\mathrm{d}}

\def\GW{\text{\tiny GW}}

\begin{document}

\title{Probing pre-Recombination Physics by the Cross-Correlation of Stochastic Gravitational Waves and CMB Anisotropies }

\author{Matteo Braglia}\email{matteo.braglia2@unibo.it}
\affiliation{Instituto de Fisica Teorica, Universidad Autonoma de Madrid, Madrid, 28049, Spain}

\affiliation{INAF/OAS Bologna, via Gobetti 101, I-40129 Bologna, Italy}

\author{Sachiko~Kuroyanagi}\email{sachiko.kuroyanagi@csic.es}
\affiliation{Instituto de Fisica Teorica, Universidad Autonoma de Madrid, Madrid, 28049, Spain}
\affiliation{Department of Physics and Astrophysics, Nagoya University, Nagoya, 464-8602, Japan}

\date{\today}
\begin{abstract}
	We study the effects of pre-recombination physics on the Stochastic Gravitational Wave Background (SGWB) anisotropies induced by the propagation of gravitons through the large-scale density perturbations and their cross-correlation with Cosmic Microwave Background (CMB) temperature and E-mode polarization ones. As examples of Early Universe extensions to the $\Lambda$CDM model, we consider popular models featuring extra relativistic degrees of freedom, a massless non-minimally coupled scalar field, and an Early Dark Energy component. 
	Assuming the detection of a SGWB, we perform a Fisher analysis to assess in a quantitative way the capability of future gravitational wave interferometers (GWIs) in conjunction with a future large-scale CMB polarization experiment to constrain such variations. Our results show that the cross-correlation of CMB and SGWB anisotropies will help tighten the constraints obtained with CMB alone, with an improvement that significantly depends on the specific model as well as the maximum angular resolution  of the GWIs, their designed sensitivity, and the amplitude $A_*$ of the monopole of the SGWB. 
\end{abstract}

\pacs{Valid PACS appear here}
\keywords{Suggested keywords}
\maketitle

\section{Introduction}

The new era of gravitational wave (GW) astronomy has the potential to drastically change our understanding of Cosmology and Fundamental Physics over a broad range of redshifts. While standard sirens \cite{Schutz:1986gp} can be used to probe the Hubble expansion up to redshifts $z\sim\mathcal{O}(10)$ and place constraints on late-time modifications to gravity \cite{Belgacem:2019pkk,Finke:2021aom}, the detection of a Stochastic Gravitational Waves Background (SGWB) is often regarded as one of the best observational windows to probe the physics operating during the early Universe. Indeed, the sensitivity of future gravitational wave interferometers (GWIs) may be enough to reconstruct the spectral shape of the SGWB \cite{Kuroyanagi:2018csn,Caprini:2019pxz,Flauger:2020qyi} making it possible to separate its cosmological and astrophysical contributions \cite{Boileau:2020rpg} and shed light on inflation, reheating, primordial black holes, phase transitions and more in general physics at energies out of the reach of future particle physics accelerators (see Refs.~\cite{Kuroyanagi:2009br,Kuroyanagi:2011fy,Kuroyanagi:2015esa,Bartolo:2016ami,Kuroyanagi:2017kfx,Sasaki:2018dmp,Fujita:2018ehq,Caldwell:2018giq,DEramo:2019tit,Caprini:2019egz,Domenech:2020kqm,Fumagalli:2020nvq,Braglia:2020taf,Fumagalli:2021cel,Calcagni:2020tvw} for an incomplete list of works).

Although most of the research on the SGWB has focused on its isotropic part, the characterization of other observables have been studied, such as higher point graviton correlation functions \cite{Bartolo:2018qqn} and their polarization \cite{Callister:2017ocg,Domcke:2019zls} and anisotropies \cite{Mentasti:2020yyd,Contaldi:2020rht,Banagiri:2021ovv}. Furthermore, another way to maximize the information that can be extracted by GW data is by cross-correlating them with those from other cosmological observations. The prime example of the power of  Multi-messenger observations is the detection of GW170817, a neutron stars merger \cite{TheLIGOScientific:2017qsa}, that revolutionized our understanding of Cosmology \cite{Ezquiaga:2018btd} and Astrophysics \cite{GBM:2017lvd,Drout:2017ijr} (see also Refs.~\cite{Mukherjee:2019wfw,Mukherjee:2019wcg,Mukherjee:2019oma,Mukherjee:2020hyn,Garoffolo:2020vtd,Alonso:2020rar,Yang:2020usq,Mukherjee:2020mha} for more recent efforts on cross-correlating GW and electromagnetic cosmological probes). 
 
In this paper, we explore the possibility of constraining the cosmological history of our Universe using joint observations of Cosmic Microwave Background (CMB) and SGWB anisotropies. Recently, a line-of-sight formalism to compute the SGWB anisotropies generated by the cosmological propagation of GWs through scalar and tensor inhomogeneities has been derived in Refs.~\cite{Contaldi:2016koz,Bartolo:2019oiq,Bartolo:2019yeu}. The formalism has been used to compute the spectra of the anisotropies induced by extra-relativistic degrees of freedom in Ref.~\cite{DallArmi:2020dar}, where it was shown that the early decoupling of gravitons makes them an excellent observable to constrain the physics of the early Universe. Their cross-correlation with CMB temperature anisotropies has instead first been considered in Refs.~\cite{Adshead:2020bji,Malhotra:2020ket} and used to forecast constraints on primordial non-Gaussian signals originated by scalar-tensor-tensor interactions.
  
Building on the intuition of Ref.~\cite{DallArmi:2020dar}, our goal is to investigate the effects of pre-recombination physics on SGWB anisotropies as well as the CMB temperature and E-mode polarization ones. In order to assess the detectability of SGWB anisotropies and the information gain resulting from the cross-correlation,  we derive the noise curves for future gravitational waves interferometers (GWIs) and perform a Fisher analysis to forecast the improvement in the error on the cosmological parameters describing the pre-recombination physics. As benchmark, we consider  three popular early time modifications to the standard model in the form of  extra relativistic degrees of freedom (as in Ref.~\cite{DallArmi:2020dar}), a massless non-minimally coupled scalar field \cite{Rossi:2019lgt,Braglia:2020iik,Ballesteros:2020sik} and Early Dark Energy (EDE) \cite{Poulin:2018cxd,Agrawal:2019lmo}. Our findings suggest that  SGWB anisotropies, and their cross-correlation with CMB ones, will be a valuable observational channel to further constrain these models.

We note that unresolved astrophysical sources of gravitational waves  can also contribute to the SGWB and to its anisotropies~\cite{Regimbau:2016ike,Cusin:2017fwz,Cusin:2018rsq,Cusin:2019jpv,Cusin:2019jhg,Bertacca:2019fnt,Pitrou:2019rjz}.
 For simplicity, in this paper, we only focus on the cosmological contribution, leaving the astrophysical one for future exploration.

Our paper is organized as follows. We review the Boltzmann formalism to compute the SGWB anisotropies induced by density perturbations in the next Section and compute the theoretical spectra of SGWB anisotropies in Section~\ref{sec:models}. In Section~\ref{sec:noise}, we compute the noise spectra adopted in our Fisher analysis, which we describe in detail in Sections~\ref{sec:Fisher}. We then present our results in Section~\ref{sec:results}  and conclude in Section \ref{sec:conclusions}.

\section{ Anisotropies of the Stochastic Gravitational Wave Background}
In this Section, we review the line-of-sight formalism that we adopt to compute the SGWB anisotropies angular spectra in the next Sections. Such Boltzmann formalism is valid in the weak field limit as long as the comoving momentum of the GW is much larger than that of the large scale perturbations~\cite{Contaldi:2016koz}, which we assume throughout this paper. For a detailed treatment, we refer to the original papers  Refs.~\cite{Bartolo:2019oiq,Bartolo:2019yeu}. 

We consider the metric in the longitudinal gauge $ds^2=a^2(\eta)\left\{-(1+2\Phi)d\eta^2+\left[(1-2\Psi)\delta_{ij}+h_{ij}\right]dx^idx^j\right\}$, where $a(\eta)$ is the scale factor and $h_{ij}$ is the transverse and traceless degrees of freedom. Following a Boltzmann approach \cite{Dodelson:2003ft}, we can define the distribution function of gravitons as $f \left( x^\mu,\,p^\mu\right)$, where $x^\mu$ and $p^\mu\equiv dx^\mu/d\lambda$ are respectively the position and momentum of the graviton and $\lambda$ is an affine parameter along its trajectory. Disregarding collisional terms for GWs (see Ref.~\cite{Bartolo:2018igk} for a discussion) and keeping only terms at first order in perturbations, the Boltzmann equation for $f$ is
\begin{equation}
\frac{\partial f}{\partial \eta}+
n^i \, \frac{\partial f}{\partial x^i} +
\left[  \frac{\partial \Psi}{\partial \eta} - n^i \, \frac{\partial \Phi}{\partial x^i} + \frac{n^i n^j}{2}   \frac{\partial  h_{ij} }{\partial \eta} \right] q \,   \frac{\partial f}{\partial q} = 0 \,,
\end{equation}
where $n^i\equiv p^i/p$ with $p\equiv \sqrt{p_i p^i}$ is the unit vector describing the direction of motion of the GW, and $q\equiv p a$ is the comoving momentum.

We decompose the distribution function into the sum of an isotropic and homogeneous part and a perturbed one as $f \left( \eta ,\, x^i ,\, q ,\, n^i \right)
\equiv  {\bar f} \left( q \right)  - q \, \frac{\partial {\bar f}}{\partial q} \, \Gamma \left( \eta ,\, x^i ,\, q ,\, n^i \right)$. The first term is related to the homogeneous energy density of GWs as 
\begin{equation}
\bar{\Omega}_\GW \left( q \right) \equiv \frac{4 \pi}{\rho_{\rm crit,0}} \, \left( \frac{q}{a_0} \right)^4 {\bar f} \left( q \right) \;,  
\end{equation}
where $0$ denotes the value today and $\rho_{\rm crit,0}=3H_0^2/(8\pi G)$ is the critical energy density with $H\equiv a'/a^2$. The Fourier transform of the perturbed term $\Gamma (\eta ,\, x^i ,\, q ,\, n^i) \equiv\int d^3k/(2\pi)^3 e^{i \vec{k}\cdot \vec{x}}\Gamma(\eta ,\, k^i ,\, q ,\, n^i)$ satisfies the following equation \cite{Contaldi:2016koz}:
\begin{equation}
\Gamma'+ i \, k \, \mu\, \Gamma = S (\eta, k^i, n^i)  \, ,
\label{Boltfirstgamma1}
\end{equation}
where a prime denotes a derivative with respect to conformal time, $\mu \equiv  (k^i/k) \cdot  n_i$, and the source term is

\begin{equation}
S  = \Psi' - i k \, \mu \, \Phi -  \frac{1}{2}n^i n^j  \, h_{ij}' \,.
\end{equation}

The function $\Gamma$ can then be expanded in spherical harmonics as $\Gamma(n^i)=\sum_\ell\sum_m\,\Gamma_{\ell m} Y_{\ell m}(n^i)$ and the statistical distribution of the harmonic coefficient is  characterized by the two point function $\left\langle \Gamma_{\ell m}   \Gamma_{\ell' m'}^*  \right\rangle \equiv \delta_{\ell \ell'} \,\, \delta_{mm'} \, \left({\widetilde C}_\ell^I(q)+{\widetilde C}_\ell^S+{\widetilde C}_\ell^h\right)$, where the three terms on the right hand side, that we assume to be uncorrelated among themselves, represent the initial contribution to the total anisotropies from the SGWB generation mechanism and the anisotropies induced by the propagation of GWs through scalar and tensor perturbations, respectively \cite{Bartolo:2019oiq,Bartolo:2019yeu}. In the following, we will discard the first and third term and set ${\widetilde C}_\ell^I \left( q \right)={\widetilde C}_\ell^h=0$ as the first one is not relevant for our purposes\footnote{Mechanisms generating non-vanishing initial anisotropies include primordial black hole formation \cite{Bartolo:2019oiq,Bartolo:2019yeu,Garcia-Bellido:2016dkw,Bartolo:2019zvb}, primordial scalar-tensor-tensor non-Gaussianities \cite{Adshead:2020bji,Malhotra:2020ket}, phase transitions in the early Universe \cite{Kumar:2021ffi} and cosmic strings \cite{Kuroyanagi:2016ugi,Jenkins:2018nty}. } and  we have checked that ${\widetilde C}_\ell^h$ is negligible compared to the scalar contribution ${\widetilde C}_\ell^S$ for observationally viable values of the GW amplitude.  Therefore, the only remaining contribution to the angular spectrum of the anisotropies is explicitly given by \cite{Bartolo:2019oiq,Bartolo:2019yeu}:

\begin{equation}
C_\ell^\Gamma\equiv{\widetilde C}_\ell^S =  4 \pi  \int \frac{dk}{k} \,  {\cal T}_\ell^{\left( S \right) \,2} \left( k ,\, \eta_0 ,\, \eta_{\rm in} \right) 
\, \mathcal{P}_{\mathcal{R}} \left( k \right)  \,,
\label{Cell-res}
\end{equation}
where  $\mathcal{P}_{\mathcal{R}}(k)$ is the primordial scalar power spectrum and the transfer function is
\begin{align}
\label{eq:transf}
&{\cal T}_\ell^{\left( S \right) } \left( k ,\, \eta_0 ,\, \eta_{\rm in} \right) \equiv T_\Phi \left( \eta_{\rm in} ,\, k \right) \, j_\ell \left( k \left( \eta_0 - \eta_{\rm in} \right) \right)\\
&+ \int_{\eta_{\rm in}}^{\eta_0} d \eta' \, 
\frac{\partial \left[ T_\Psi \left( \eta ,\, k \right) +  T_\Phi \left( \eta ,\, k \right) \right] }{\partial \eta} \, 
j_\ell \left( k \left( \eta_0 - \eta \right) \right),\nonumber
\end{align} 
with $T_\Phi$ and $T_\Psi$ being the transfer functions for the Newtonian potentials and $j_\ell (x)$ the Bessel functions.

\begin{figure*}
\includegraphics[width=.49\columnwidth]{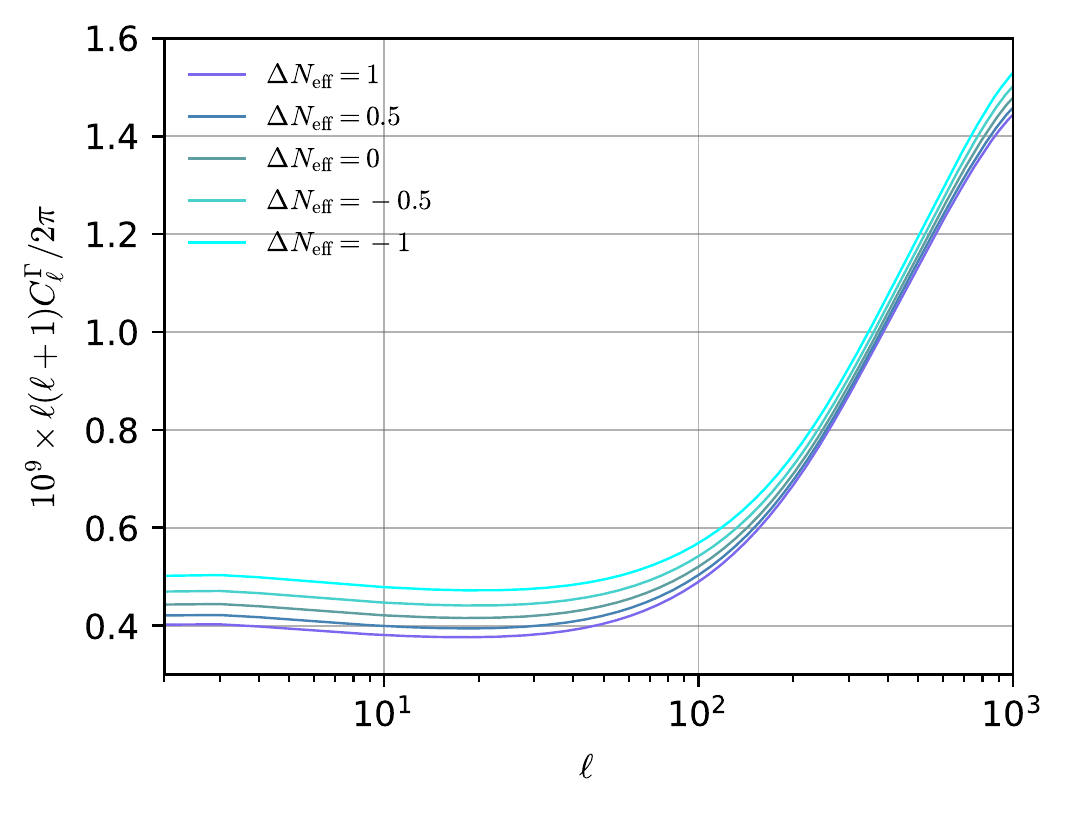}
\includegraphics[width=.49\columnwidth]{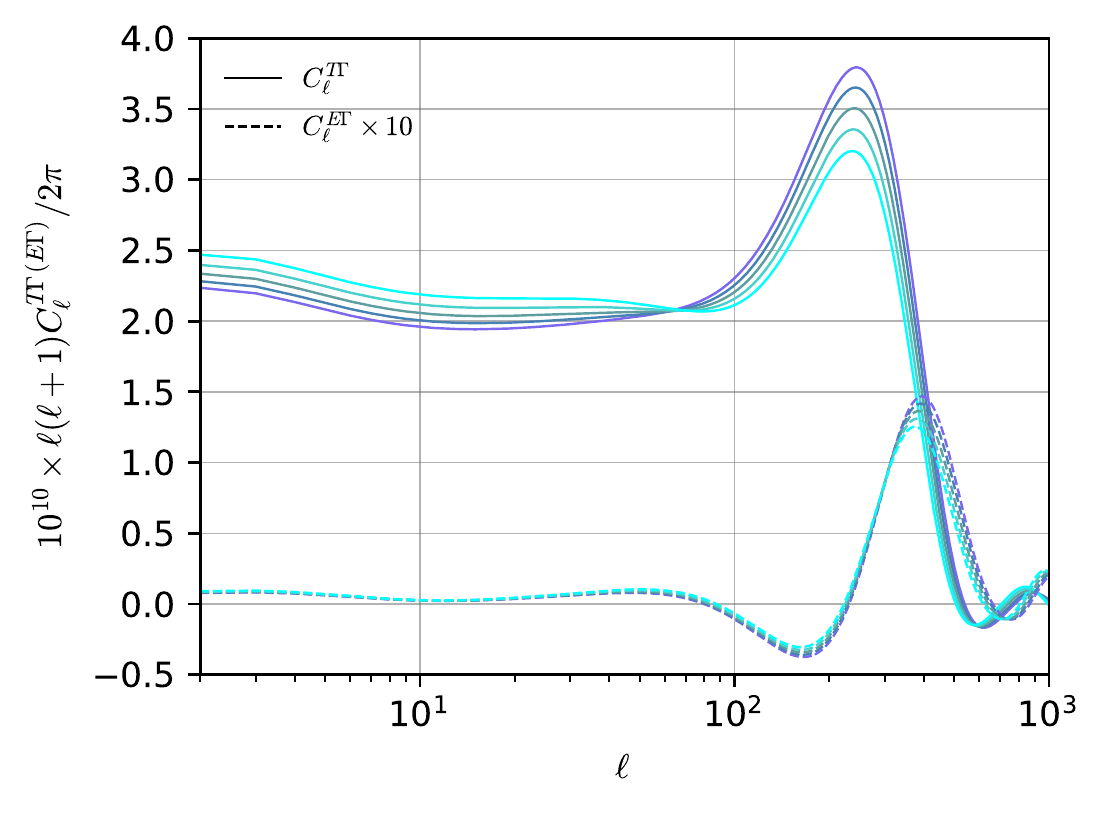}
\caption{\label{fig:ClNeff} Spectra of SGWB anisotropies in the model with extra relativistic degrees of freedom. The angular power spectra are shown for [Left] the variable $\Gamma$ and [right] their cross-correlation with the CMB T (solid) and E (dashed) spectra. The parameter $\Delta N_{\rm eff}=N_{\rm eff}-3.046$  is varied according to the legend. }
\end{figure*}

As for CMB photons, the scalar potentials induce a Sachs-Wolfe (SW) effect which dominates at large scales, i.e. small $\ell$'s, given by the first term in  ${\cal T}_\ell^{\left( S \right) }$, and an integrated SW (ISW) effect, which depends on the variation of the scalar potentials integrated along the line of sight \cite{Contaldi:2016koz,Bartolo:2019oiq,Bartolo:2019yeu}. The crucial difference with the CMB, however, is the absence of the visibility function $g(\tau)=-(d\tau/dt) e^{-\tau}$, where $\tau$ is the optical depth, which effectively sets $\eta_i\simeq\eta_*$ for CMB photons, where $\eta_*$ is around the last scattering surface \cite{DallArmi:2020dar}.

Although the variable $\Gamma$ is the most natural to adopt in the Boltzmann approach, it is not the mostly used when it comes to studying the noise for the angular spectra of the SGWB anisotropies. In particular, we will compute the noise spectra with schNell code \cite{Alonso:2020rar}, which is optimized to compute the anisotropies of the energy density of GWs $\Omega_{\rm GW}(f,\,n^i)$ itself. The latter is related to $\Gamma$ by \cite{Bartolo:2019oiq,Bartolo:2019yeu}
\begin{equation}
\label{eq:Omega-Gamma}
\Omega_{\rm GW}(f,\,n^i)\equiv\bar{\Omega}_{\rm GW}(f)\left[1+\left(4-\alpha\right)\Gamma\right],
\end{equation}
where, for simplicity, we restrict to a power-law monopole $\Omega_{\rm GW}(f)$ of the form
\begin{equation}
\bar{\Omega}_{\rm GW}(f)=A_*\left(\frac{f}{f_*}\right)^\alpha\,,
\end{equation}
with $\alpha$ being the spectral index and $f_*$ being a reference frequency such that $A_*=\bar{\Omega}_{\rm GW}(f_*)$. For a more general form of $\bar{\Omega}_{\rm GW}$, the formulae that we provide below are still valid after the substitution  $\alpha\mapsto\partial\ln\bar{\Omega}_{\rm GW}(f)/\partial \ln f$.

Using these relations and expanding $\Omega_{\rm GW}$ in spherical harmonics as $\Omega_{\rm GW}(f,\,n^i)\equiv\bar{\Omega}_{\rm GW}(f)\left(1+\delta_{\rm GW}(f,\,n^i)\right)$, we can relate the angular power spectrum of $\Omega_{\rm GW}$ to the one of $\Gamma$ as:

\begin{equation}
\label{eq:CellOmega}
C_\ell^\Omega(f)=A_*^2\left(\frac{f}{f_*}\right)^{2\alpha}(4-\alpha)^2C_\ell^\Gamma.
\end{equation}
Note that the $C_\ell^\Omega$s are frequency dependent. In this paper, for simplicity, we will always compute them at the reference frequency $f_*$.

As anticipated in the Introduction, we will also be interested in computing the cross-correlation of the SGWB anisotropies with CMB ones. We define such cross-correlation as:
\begin{equation}
     C_\ell^{\Omega\,X}   =  4 \pi \, \int \frac{d k}{k} \,  \left[ {\cal T}_\ell^S \left( k\right) \Delta^X(k)\right] 
\,  \mathcal{P}_{\mathcal{R}} \left( k \right),
\end{equation}
where $X=\{T,\, E\}$ and $\Delta^X(k)$ is the transfer function for CMB photons.

In order to compute the spectra of the SGWB anisotropies, we have modified the publicly available code {\tt CLASS}\footnote{\href{https://github.com/lesgourg/class\_public}{https://github.com/lesgourg/class\_public}} 
\cite{Lesgourgues:2011re,Blas:2011rf}.

\section{Selected cosmological models and theoretical spectra}
\label{sec:models}
We now present the theoretical spectra for the SGWB anisotropies and their cross-correlation with the CMB ones. For each of our benchmark models for early time modifications, we first review the main features and then numerically explore how they affect the angular spectra of the anisotropies.

\subsection{Extra relativistic degrees of freedom}
\label{sec:DeltaNeff}

We start by considering the contribution of extra relativistic species.  In this context, modifications to the expansion history of the early Universe are often enclosed in the effective number of relativistic species $N_{\rm eff}$, defined as
\begin{equation}
     \label{eq:Neff}
     \rho_r=\left[1+\frac{7}{8}\left(\frac{4}{11}\right)^{\frac{4}{3}}N_{\rm eff}\right]\rho_\gamma,
\end{equation}
where $\rho_r$ is the radiation energy density and $\rho_\gamma$ is the photon energy density. For the Standard Model of particle physics, there are three species of active neutrinos corresponding to $\Delta N_{\rm eff} \equiv N_{\rm eff}-3.046 = 0$, where the small correction  $N_{\rm eff}-3=0.046$ accounts for the fact that neutrino decoupling is immediately followed by $e^+\,e^-$ annihilation, see e.g. Ref.~\cite{Lesgourgues:2018ncw}.

As mentioned in the Introduction, the imprints of relativistic degrees of freedom on the spectra of the SGWB anisotropies were first studied in Ref.~\cite{DallArmi:2020dar}, to which the reader is referred for more details. The formalism of Ref.~\cite{DallArmi:2020dar} is a bit more general than ours, as the authors analyze the response of  SGWB anisotropies to a change in the fractional energy density of decoupled relativistic species at the time of graviton decoupling, i.e. $\eta_i$, defined as $f_{\rm dec}(\eta_i)=g_*^{\rm dec}(T_i)/g_*(T_i)$, where $g_*$ and $g_*^{\rm dec}$ are the relativistic degrees of freedom in the standard model and the ones of decoupled particles, respectively. The latter parameterization therefore allows to include decoupled relativistic particles at early times that are no longer relativistic today and therefore do not contribute to $N_{\rm eff}$. In our paper, for simplicity, we restrict to extra-relativistic species that are still relativistic today and are already decoupled at the end of inflation, as required to impact the low-$\ell$ anisotropies of the SGWB \cite{DallArmi:2020dar}.

As can be seen from Fig.~\ref{fig:ClNeff}, a variation in $N_{\rm eff}$ affects high multipoles, since the redshift of the matter-radiation equality, or equivalently the comoving size sound horizon $r_s$, is changed by the extra contribution to the Hubble rate. However, as noticed in Ref.~\cite{DallArmi:2020dar}, since gravitons decouple earlier than CMB photons, the Sachs-Wolfe plateau is also affected, unlike in the CMB spectra.

\subsection{Non-Minimally coupled scalar field}
\label{sec:CC}

\begin{figure}
    \begin{center}
    \includegraphics[width=\columnwidth]{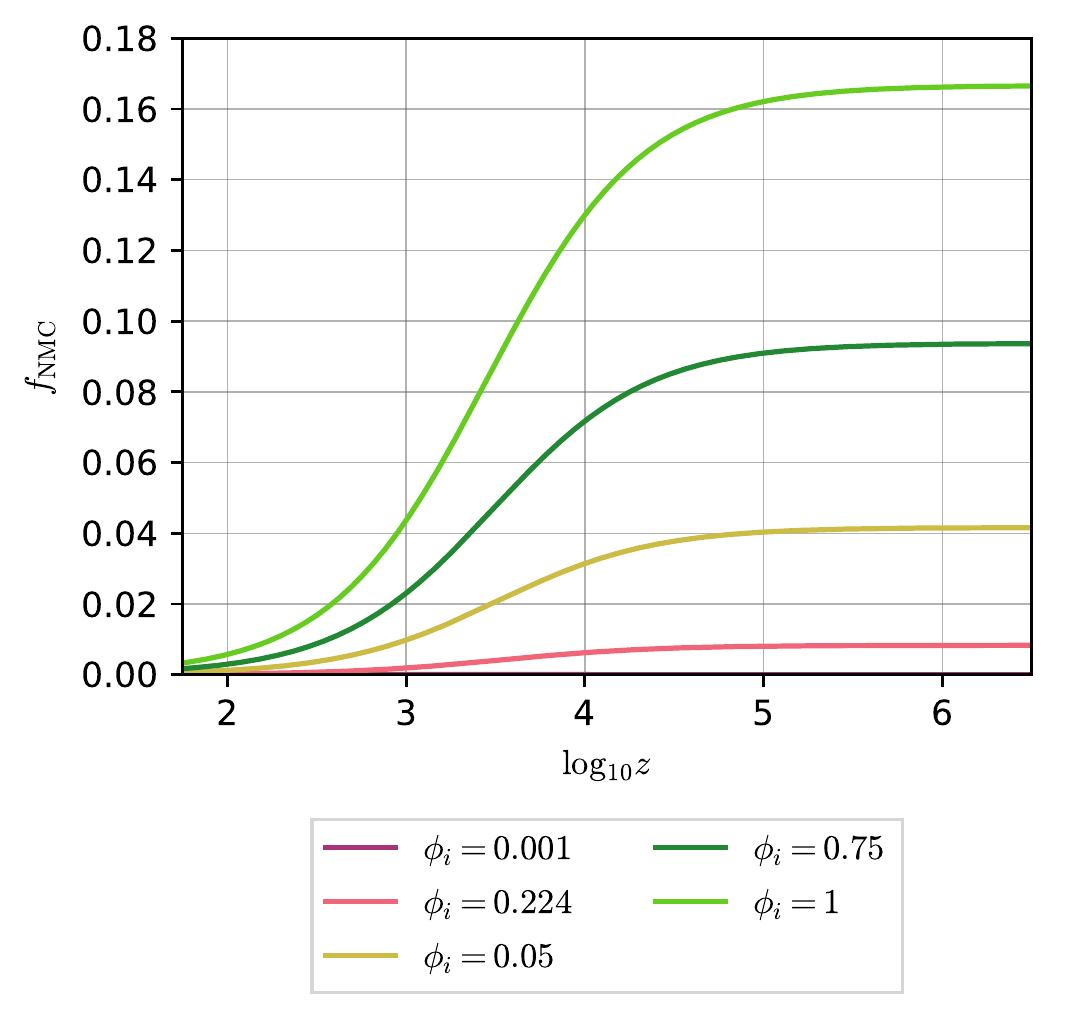}
    \end{center}
    \caption{\label{fig:NMC} Energy injection in the NMC model. The initial condition of the scalar field $\phi_i$ (in Planckian units) is varied according to the legend and we fix $\xi=-1/6$.}
\end{figure}

\begin{figure*}
\includegraphics[width=.49\columnwidth]{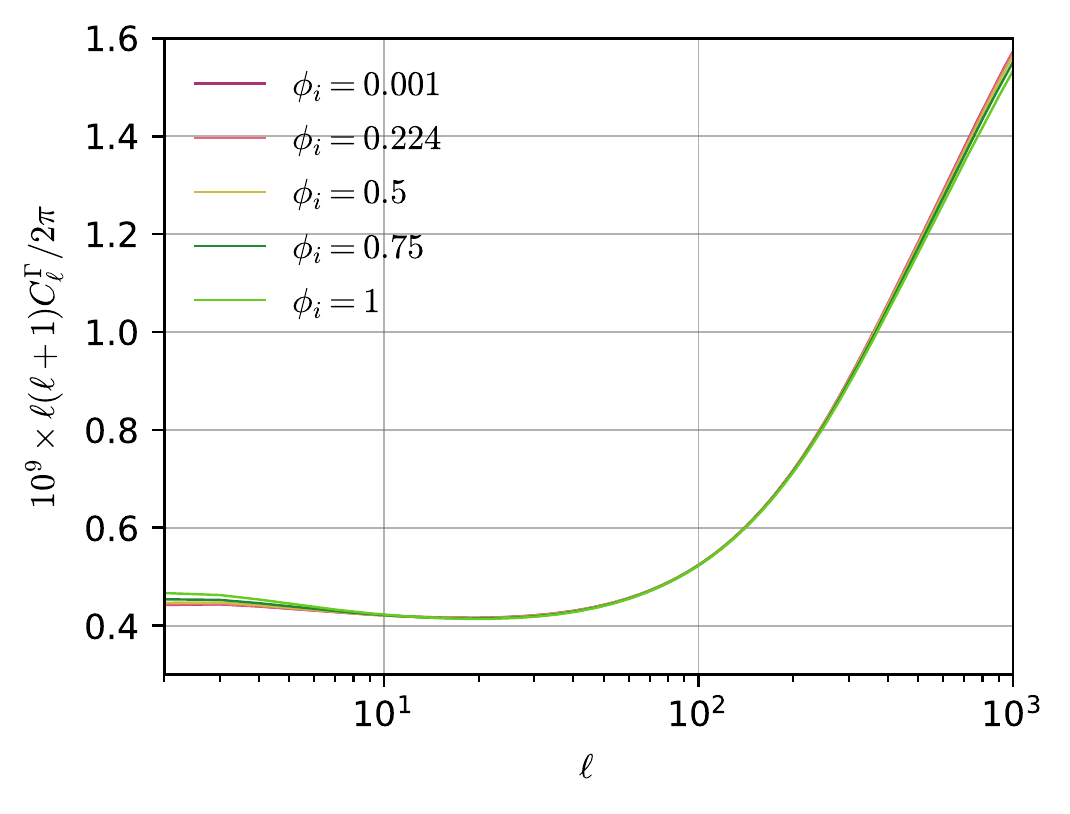}
\includegraphics[width=.49\columnwidth]{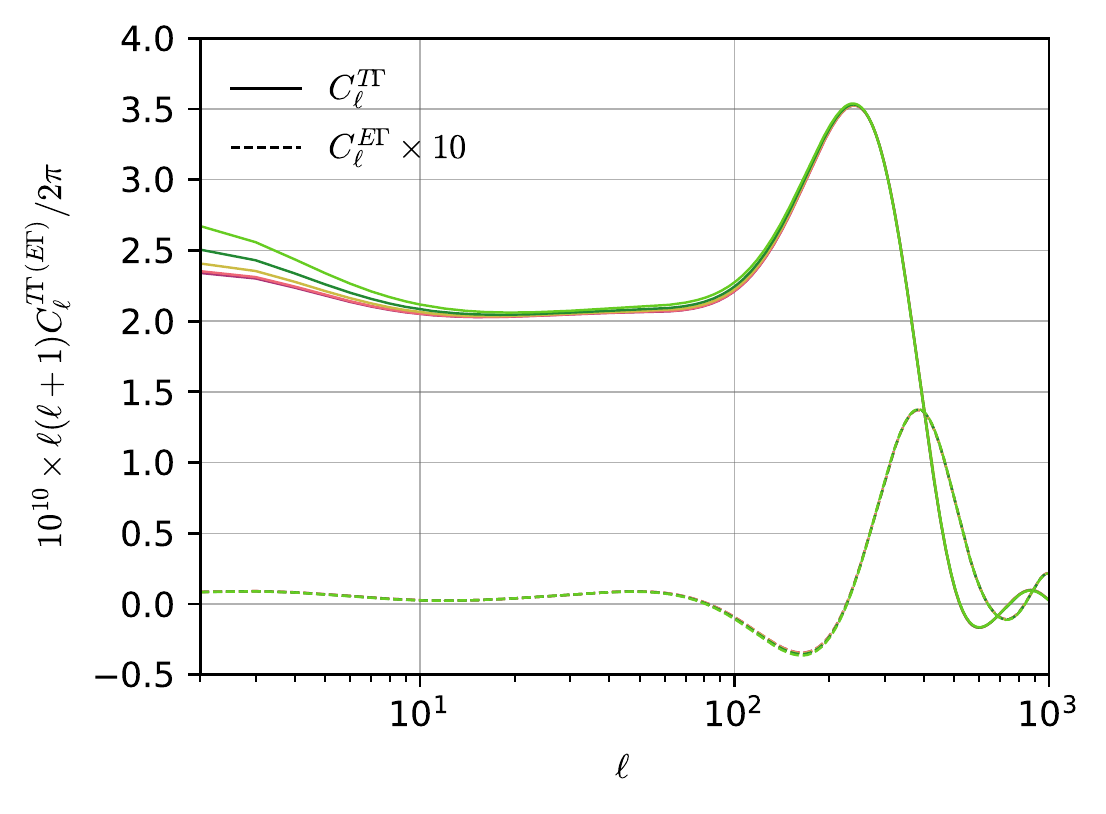}
\caption{\label{fig:ClCC} Spectra of SGWB anisotropies in the NMC model. The angular power spectra are shown for [Left] the variable $\Gamma$ and [right] their cross-correlation with the CMB T (solid) and E (dashed) spectra. The initial condition of the scalar field $\phi_i$ (in Planckian units) is varied according to the legend and we fix $\xi=-1/6$.}
\end{figure*}

As a second example, we consider a Non-Minimally coupled (NMC) scalar field $\phi$ that modifies gravity before and around recombination \cite{Rossi:2019lgt,Braglia:2020iik,Ballesteros:2020sik}, described by the following Lagrangian:
\begin{equation}
\label{eq:modelCC}
    S = \int \dd^{4}x \sqrt{-g} \left[ \frac{F(\phi)}{2}R 
    - \frac{(\partial\phi)^2}{2} -\Lambda\right]+ S_m \,,
\end{equation}
where $F(\phi) = M_{\rm pl}^2+\xi\phi^2$, $R$ is the Ricci scalar, and $S_m$ is the action for matter fields. In this model, the scalar field is frozen deep into the radiation era and starts to move around the matter-radiation era driven by its coupling to pressureless matter at the level of the equations of motion.

For negative values of the coupling $\xi<0$, this model modifies the expansion history of the Universe by contributing with a nearly constant energy fraction before recombination, after which it redshifts away as fast or faster than radiation, depending on the magnitude of $\xi$ \cite{Braglia:2020iik,Ballesteros:2020sik}. We show the dependence of this contribution, quantified by the quantity $f_{\rm NMC}\equiv\rho_{\rm NMC}/\rho_{\rm crit}$, where $\rho_{\rm NMC}$ is the effective energy density of the scalar field \cite{Gannouji:2006jm}, in Fig.~\ref{fig:NMC}. As can be seen, the magnitude of $f_{\rm NMC}$ increases with the initial condition of the scalar field $\phi_i$.

Although the background expansion history resembles the one of the $\Delta N_{\rm eff}$ model introduced before, the behavior of the perturbations is different \cite{Rossi:2019lgt} resulting in distinct cosmological predictions \cite{Ballesteros:2020sik}. We show the theoretical spectra for the NMC model, focusing for simplicity on the so-called Conformally Coupled case $\xi=-1/6$ in Fig.~\ref{fig:ClCC}.

\subsection{Early Dark Energy}
\label{sec:EDE}

\begin{figure}
    \centering
    \includegraphics[width=\columnwidth]{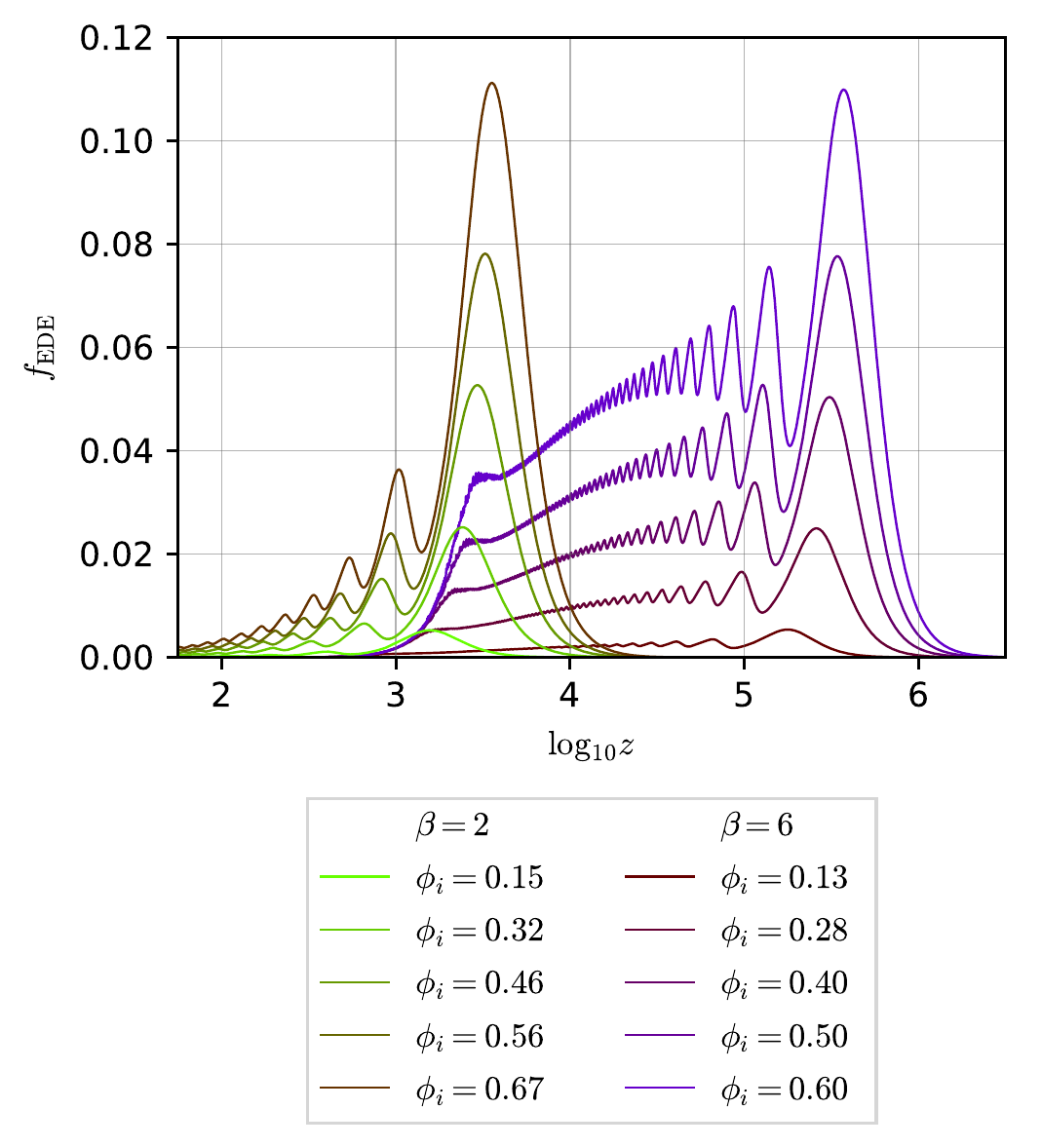}
    \caption{\label{fig:fEDE} Energy injection in the RnR model. The parameter $\phi_i$ is varied according to the legend (in Planckian unit). We plot examples for the redshift of energy injection $z_c=3160$ and $z_c=7.3\times10^5$, corresponding to $\beta=2$ and $\beta=6$, respectively.}
    \label{fig:fEDE}
\end{figure}

\begin{figure*}
\includegraphics[width=.49\columnwidth]{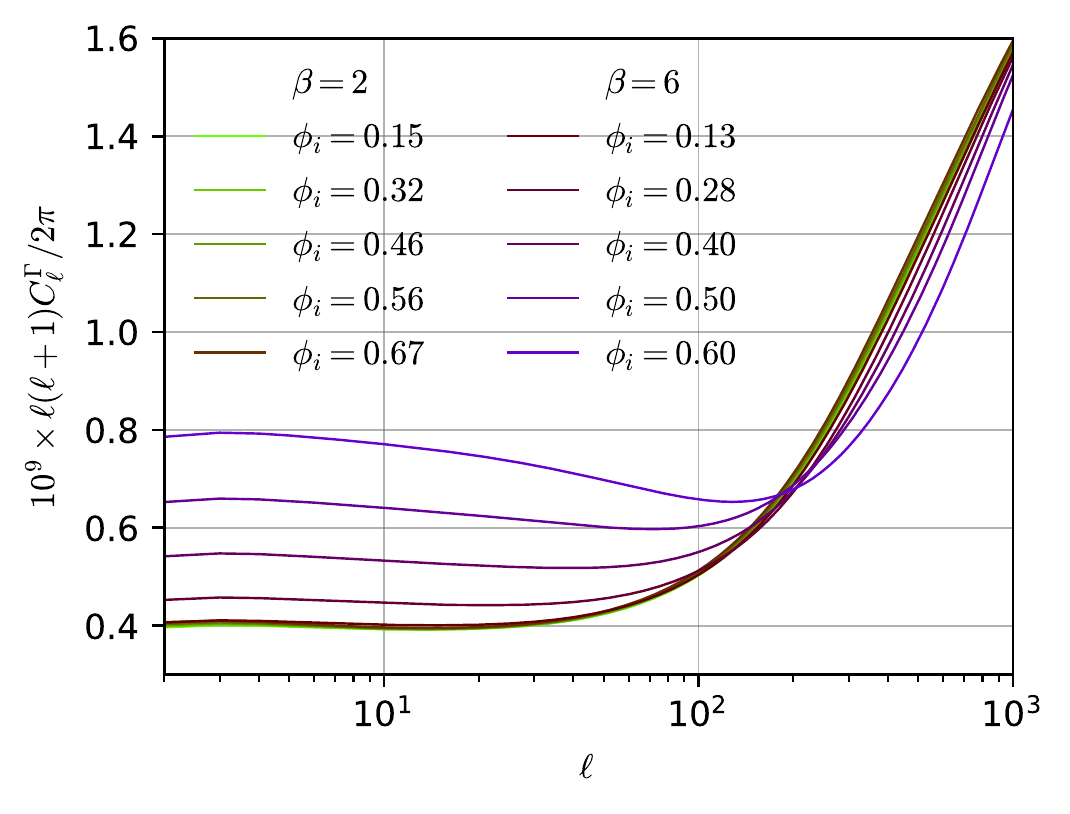}
\includegraphics[width=.5\columnwidth]{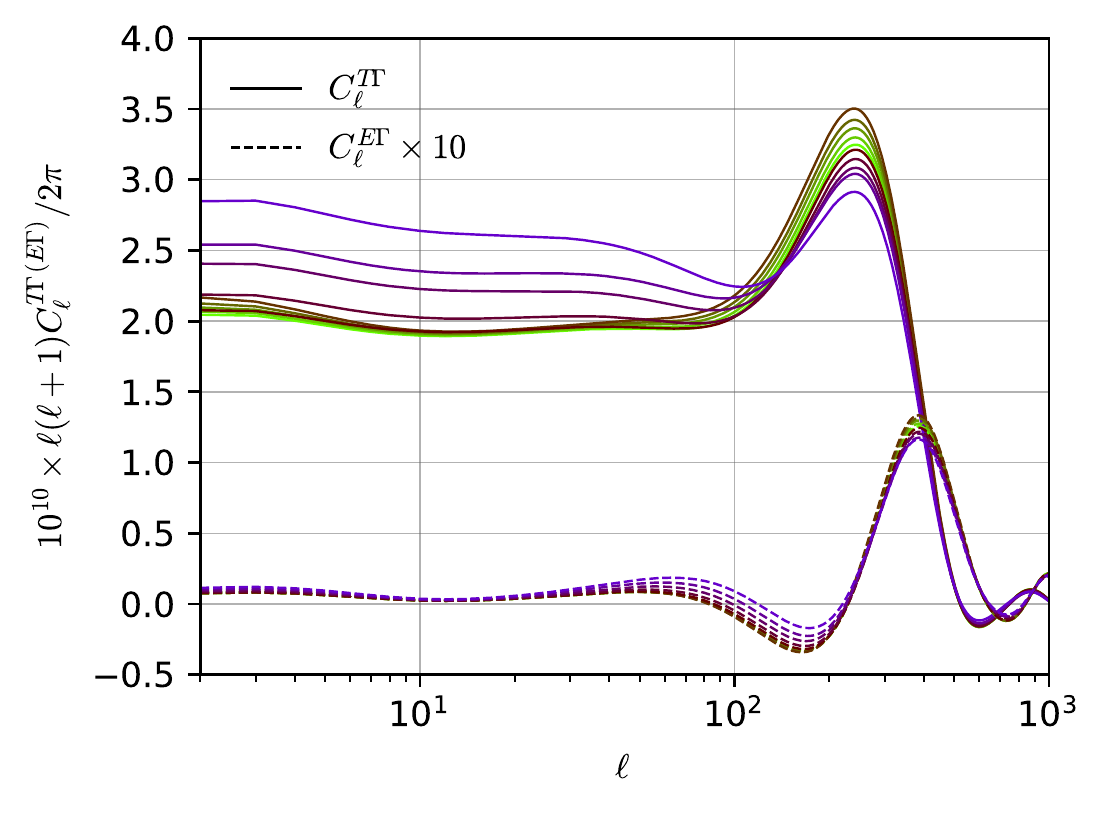}
\caption{\label{fig:ClNE} Spectra of SGWB anisotropies in the EDE model. The angular power spectra are shown for [Left] the variable $\Gamma$ and [right] their cross-correlation with the CMB T (solid) and E (dashed) spectra. The parameter $\phi_i$ is varied according to the legend (in Planckian unit). We plot spectra for $z_c=3160$ and $z_c=7.3\times10^5$, corresponding to $\beta=2$ and $\beta=6$, respectively.  }
\end{figure*}

As a last example, we consider the case of Early Dark Energy (EDE). This class of models has recently become popular as a solution to the $H_0$ tension \cite{Verde:2019ivm}. Here, we consider the so called Rock n Roll (RnR) model of Ref.~\cite{Agrawal:2019lmo} as representative of EDE (see \cite{Poulin:2018cxd,Agrawal:2019lmo,Niedermann:2019olb,Berghaus:2019cls,Sakstein:2019fmf,Lin:2019qug,Ye:2020btb,Braglia:2020iik,Braglia:2020bym,Gonzalez:2020fdy,Braglia:2020auw} for an incomplete list of EDE models). The Lagrangian is that of a minimally coupled, canonical scalar field $\phi$:
\begin{equation}
\label{eq:modelEDE}
    S = \int \dd^{4}x \sqrt{-g} \left[ \frac{R}{2} 
    - \frac{(\partial\phi)^2}{2} -\Lambda-V(\phi)\right]+ S_m \,,
\end{equation}
where $V(\phi)$ is a quartic potential of the form $V(\phi)=\lambda\phi^{2 n}/2n$. We take $n=2$ and parameterize the dimensionless constant $\lambda$ as $\lambda=10^{2 \beta}/(3.516\times10^{109})$, where $3.516\times10^{109}$ is the \emph{numerical} value of $M_{\rm pl}^4$ in ${\rm eV}^4$.

The cosmological dynamics of the scalar field can be summarized as follows. Deep in the radiation era, the scalar field is frozen by the Hubble friction to its initial value. At these times, the kinetic energy of the scalar field is essentially zero. Therefore, the energy of the scalar field is subdominant with respect to that of radiation, and it has an equation of state $w_{\phi}=-1$, hence the name \emph{Early Dark Energy}.  When the effective mass of $\phi$ becomes comparable to the Hubble parameter, i.e. $d^2 V/d\phi^2 \simeq H^2$, it quickly rolls down the potential and eventually oscillates around its minimum at $\phi=0$, with a cycle-averaged equation of state given by \cite{Turner:1983he}:
\begin{equation}
    w_\phi=\frac{n-1}{n+1}.
\end{equation}

For the specific example of the quartic potential with $n=2$, we get $w_\phi=1/3$ and the scalar field energy density redshifts away as fast as radiation. This is a crucial feature of EDE models that makes them a perfect candidate to solve the $H_0$ tension since these models sizeably modify the Hubble expansion only in a narrow redshift range around recombination and leave unaffected the very early and the late dynamics of the Universe.

However, if the scalar field is massive enough so that it starts to move early during the radiation era, the radiation-like behavior of the envelope of the oscillations is not necessarily negligible. These two different regimes are shown in Fig.~\ref{fig:fEDE}, where we plot the EDE fractional contribution $f_{\rm EDE}=8\pi G \rho_\phi/3 H^2$ to the total density of the Universe for different masses of the scalar field. Note that we work directly with the parameters from the Lagrangian and follow the conventions of Ref.~\cite{Braglia:2020auw}. Another possibility is to trade the model parameters to $z_c$ and $f_{\rm EDE}$, which are respectively the redshift when the scalar field starts to move and the amount of energy injected into the cosmic fluid at $z_c$.

We show the effect of varying the fraction of energy injected into the cosmic fluid in Fig.~\ref{fig:ClNE}. First, we fix the value of $z_c=3160$, i.e., the best-fit value found in Ref.~\cite{Agrawal:2019lmo}. For this value, the model was originally found to sizably alleviate the $H_0$ tension (see Ref.~\cite{Braglia:2020auw} for analysis with latest data). In this case, only high multipoles are affected, and we see no observable effect on the spectra of the SGWB anisotropies at the large scales.

Then, we fix $z_c$ to a much larger value $z_c=7.3\times10^5$ and plot again the variation of the spectra with $f_{\rm EDE}$, using different colors. Now the EDE component has almost completely diluted away at the redshift of recombination, and the background effect for $z<z_c$ is similar to effectively adding new relativistic species for redshift smaller than $z_c$. Since gravitons decouple much earlier than recombination, their anisotropies are affected by the variation of the Newtonian potential induced by the presence of EDE, through the integrated Sachs-Wolfe effect in Eq.~\eqref{eq:transf}. Note, however, the situation is somewhat different from the case of extra-relativistic degrees of freedom. First of all, the background evolution is not the same. Second, unlike extra-relativistic species, minimally coupled scalar fields such as EDE do not contribute to the anisotropic stress perturbations. As a result, the SGWB spectra and their cross-correlation with CMB ones are different from those in Fig.~\ref{fig:ClNeff}.

\section{Noise angular spectra}
\label{sec:noise}
In order to forecast the capabilities of different GWIs to constrain the cosmological models that we have just presented, the noise angular power spectra $N_\ell$ for each GWI have to be specified in the Fisher analysis of the next Section. A code for the efficient computation of $N_\ell$, ready to be used for forecasting, has been recently presented in Ref.~\cite{Alonso:2020rar} and made publicly available at the link in this footnote\footnote{\href{https://github.com/damonge/schNell/branches}{https://github.com/damonge/schNell/branches}}. The code, called schNell, was used in \cite{Alonso:2020rar} to compute the noise angular power spectra of several GWIs including LISA, LIGO and some combinations of ground-based interferometers. We modified it to include also other experiments, as discussed below. Here, we only wish to present the SGWB anisotropies noise spectra, referring the interested reader to Ref.~\cite{Alonso:2020rar} for all the details regarding its computation and the implementation in schNell.

\begin{figure*}
\includegraphics[width=\columnwidth]{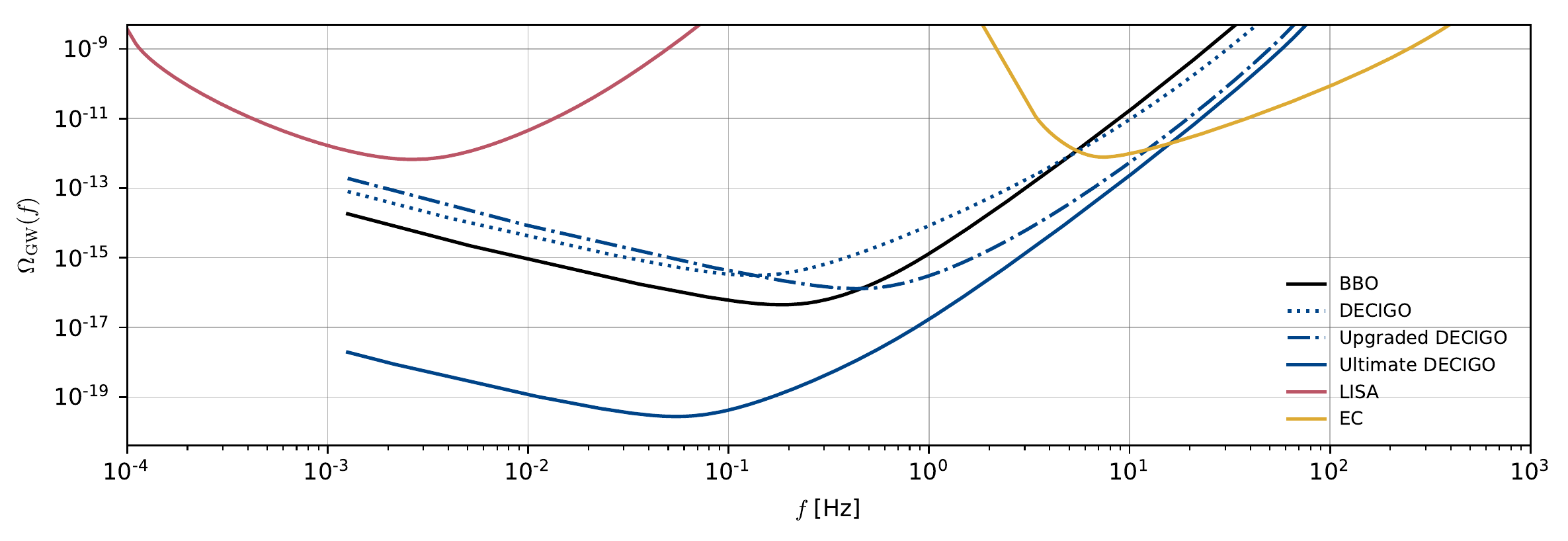}
\caption{\label{fig:PLI} We plot the power-law integrated sensitivity curves with SNR$=5$ for the experiments mentioned in the main text. }
\end{figure*}

In this paper, we consider several planned GWIs. As for future space-based detectors, we examine LISA, BBO, and  DECIGO. As for ground-based detectors, we only consider the example, presented in Ref.~\cite{Alonso:2020rar}, of a cross-correlation between the Einstein Telescope and the Cosmic Explorer, which we refer to as EC. The sensitivity of other current or planned GWIs is not enough to measure the tiny SGWB anisotropies studied in this paper. We show the power-law integrated sensitivity (PLS) curves \cite{Thrane:2013oya} in Fig.~\ref{fig:PLI}. The meaning of these curves is that every power-law $\Omega_{\rm GW}$ that crosses them will be measured with a given signal-to-noise ratio (SNR) \cite{Thrane:2013oya}, which we take to be ${\rm SNR}=5$ in Fig.~\ref{fig:PLI}.

The overlap reduction function and the noise power spectral density (PSD) for each GWI have to be specified in schNell to compute the noise spectra.  For LISA and EC, their choice is specified in Ref.~\cite{Alonso:2020rar}. For BBO and DECIGO, we use the noise PSD and the overlap reduction functions given in Ref.~\cite{Kuroyanagi:2014qza} (see also \cite{Kudoh:2005as}). Note that, for DECIGO, we plot three examples. In addition to the standard DECIGO configuration, we plot its 'Upgraded' configuration, whose sensitivity is improved by about a factor of 3, and the 'Ultimate' configuration, which is the most optimistic version of DECIGO with the sensitivity only limited by quantum noise. Since the PLS curves fore DECIGO and upgraded DECIGO are not significantly different from the one of BBO, in the rest of this paper, we will only consider ultimate DECIGO, which has a far better sensitivity than BBO. For space-based GWIs, also the orbital motion has to be specified. For BBO and DECIGO, whose specifications are not firmly established yet, we describe their coordinates using the default motion of LISA implemented in schNell, which is the one discussed in Ref.~\cite{Rubbo:2003ap}. Note also that, while we consider the hexagram configuration of BBO and DECIGO,  the proposal of an improved version, where the main hexagram is supplemented with two additional LISA-like interferometers, has also been discussed \cite{Yagi:2011wg}.

\begin{figure}
\includegraphics[width=\columnwidth]{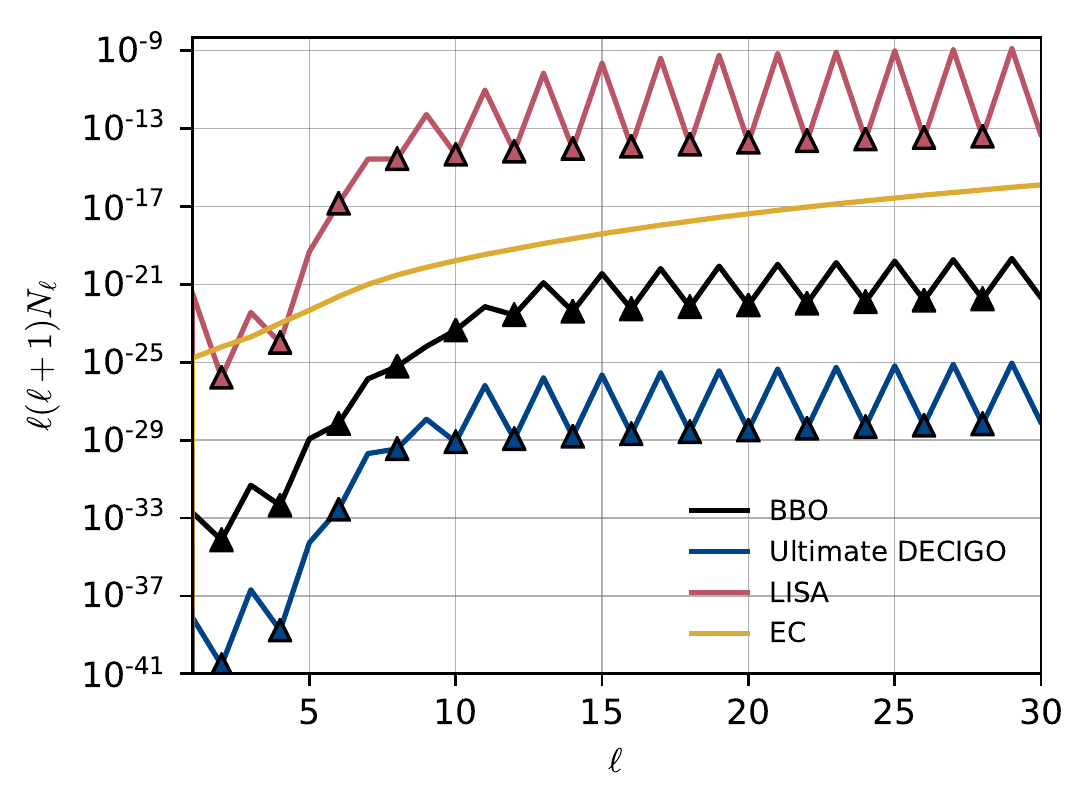}
\caption{\label{fig:Nell} We plot the noise angular power spectrum $N_\ell$ for the GWIs considered in our analysis. For LISA, BBO, and Ultimate DECIGO, we mark even multipoles with small triangles to highlight the different sensitivity to odd multipoles. All noise spectra are computed at the peak frequency of each experiment.  }
\end{figure}

We show the noise angular power spectra for our benchmark GWIs in Fig.~\ref{fig:Nell}. In producing the noise curves,  we considered a scale-invariant SGWB with $\alpha=0$. We stress that the noise curves refer to the quantity $C_\ell^\Omega$, not $C_\ell^\Gamma$. The two quantities are related by Eq.~\eqref{eq:CellOmega}. As can be seen from Fig.~\ref{fig:Nell}, due to the symmetry properties of the overlap reduction functions of space-based GWIs, they are most sensitive to even multipoles. However, for BBO and DECIGO, the difference between the sensitivity of even and odd multipoles is slightly reduced due to their hexagram configuration \cite{Kudoh:2005as}.

\section{Fisher methodology }
\label{sec:Fisher}
Having discussed the noise spectra, we go on to describe the details of our Fisher analysis. The matrix $\mathbf{C}_\ell$ containing the information from the anisotropies of the CMB and SGWB and their cross-correlation spectra has the following form: 
 \begin{eqnarray}
 \label{eq:Cl}
 	\centering
 	\left( \begin{array}{ccc}C_\ell^{TT} + N_\ell^{TT} & C_\ell^{TE}& C_\ell^{T-GW}  \\C_\ell^{TE}&C_\ell^{EE}+N_\ell^{EE}&C_\ell^{E-GW}\\C_\ell^{T-GW}& C_\ell^{E-GW} & C_\ell^{GW}+N_\ell^{GW} \end{array}\right).
	\label{eq:cov_definition}
\end{eqnarray}
Then, the Fisher matrix is simply given by 
\begin{align}
    F_{ij}&=\sum_{\ell=2}^{\ell_{\rm max}^{\rm GW}}\frac{2\ell+1}{2}f_{\rm sky}{\rm Tr}\left(\mathbf{C}_\ell^{-1}\frac{\partial \mathbf{C}_\ell}{\partial p_i}\mathbf{C}_\ell^{-1}\frac{\partial \mathbf{C}_\ell}{\partial p_j}\right)\nonumber\\
    &+\sum_{\ell=\ell_{\rm max}^{\rm GW}+1}^{\ell_{\rm max}}\frac{2\ell+1}{2}f_{\rm sky}{\rm Tr}\left(\mathbf{c}_\ell^{-1}\frac{\partial \mathbf{c}_\ell}{\partial p_i}\mathbf{c}_\ell^{-1}\frac{\partial \mathbf{c}_\ell}{\partial p_j}\right),
\end{align}
where $\mathbf{c}_\ell'$ is the matrix obtained removing the third row and column from Eq.~\eqref{eq:Cl} and $\ell_{\rm max}^{\rm GW}$ is the angular resolution of the GWIs.  Assuming the likelihood functions for the parameters $p_j$ to be Gaussian, the most optimistic errors on $p_i$ can be estimated with the Cramer-Rao bound $\sigma_i^2=F_{ii}^{-1}$. We assume a conservative sky fraction of $f_{\rm sky}=0.7$ for both CMB and GW experiments.

As for the noise spectra, $N_\ell^{\rm GW}$ is computed as outlined in the previous Section, whereas for the CMB, we use white noise spectra given as
 \begin{equation}
 	\centering
		N^{ X X' }_\ell = s^{\, 2} \exp \left(\ell(\ell+1) \frac{\theta^{\ 2}_{\textsc{fwhm}}}{8\log2}\right)\,,
	\label{eq:beamnoise}
\end{equation}
where $s$ is the total intensity instrumental noise in $\mu$K-radians and $X X'=\{TT,\,EE\}$. 
As an example of a future large-scale CMB polarization experiment we use the LiteBIRD experiment \cite{Hazumi:2019lys} for which we use the noise specifications given in Table (3.1) of \cite{Hazra:2018eib} and fix  $\ell_{\rm max}=1350$. Here an important comment is in order. We note that all the models we consider in this paper introduce new physics before and/or around the time of recombination and, therefore, modify the acoustic peak structure of the CMB spectra. As such, the maximum multipole  $\ell_{\rm max}=1350$ mapped by LiteBIRD would not be competitive to constrain these models, compared to the large multipoles already mapped by Planck. The optimal one would be the proposed CMB S4 experiment, which can map the CMB spectra over the multipole range of $30\leq\ell\leq3000$. However, it is not possible to exploit the cross-correlation of CMB and SGWB anisotropies for this experiment, as GWIs are sensitive to small multipoles (see below). In addition, to perform a robust forecast including such large multipoles, the spectra of CMB lensing potentials also have to be included, which goes beyond the scopes of our paper. For these reasons, we choose to consider LiteBIRD, which will provide the best measurement of the CMB polarization spectra at small multipoles, and leave the task of a full Fisher analysis considering CMB S4 experiments and including lensing spectra for future work.

We forecast the capability of SGWB and CMB anisotroies cross-correlation measurements to constrain the models presented in Section~\ref{sec:DeltaNeff}-\ref{sec:EDE} using our Fisher formalism, and compare them with the case of CMB only to show the improvement of adding GW data. For simplicity, we focus on a scale-invariant SGWB with $\alpha=0$, but our results are not significantly affected by this specific choice. For each experiment, we fix $f_*$ to the frequency where the sensitivity is maximum, and vary $A_*$ in the range $A_*\in[10^{-11},\,10^{-9},\,10^{-7}]$ and the angular resolution $\ell_{\rm max}^{\rm GW}\in[2,\, 30]$. We note that the LIGO O3 run already sets a bound $A_*\leq5.8\times10^{-9}$ at 95\%CL for a flat spectrum~\cite{KAGRA:2021kbb}. Therefore, the case of $A_*=10^{-7}$ for EC, whose sensitivity frequency range overlaps with that of LIGO-Virgo should be interpreted only as an illustrative example. Since, except for Pulsar Timing Arrays which are not considered in this work, no limits on $A_*$ are currently available at frequencies smaller than LIGO-Virgo ones, the possibility that the spectrum is flat at those scales cannot be completely disregarded unless one extrapolates the LIGO-Virgo bound to those scales.

\section{Results}
\label{sec:results}
We now present the results of our Fisher analysis for each model. Since all these models have been recently proposed as solutions to the tension between local and early-time measurements of the Hubble parameter, as fiducial cosmologies, we considered the best-fit parameter values that best ease the tension. We provide the fiducial parameters in each of the following subsections. In our analysis, we vary the cosmological parameters (including the six standard ones) altogether, although, to keep the discussion simpler, we only focus it on the extra parameter(s) characterizing each model.

\begin{figure*}
\includegraphics[width=\columnwidth]{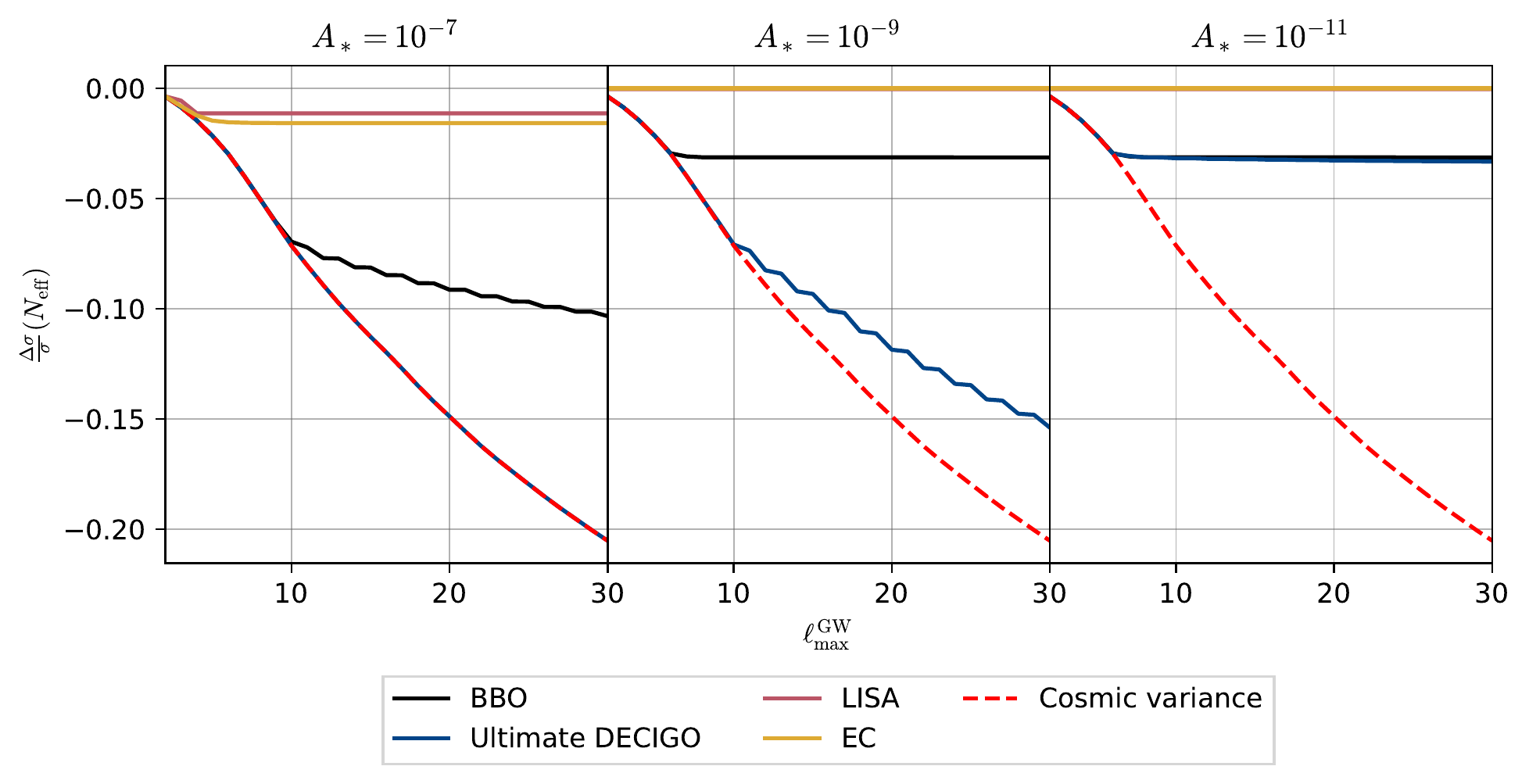}
\caption{\label{fig:sigmaNeff} Results for the $\Delta N_{\rm eff}$ model. We plot the relative improvement $\Delta\sigma/\sigma\equiv (\sigma_{\rm CMB + SGWB}-\sigma_{\rm CMB})/\sigma_{\rm CMB}$ of the error on $N_{\rm eff}$, plotted by changing the angular resolution of the GW experiments up to $\ell_{\rm max}^{\rm GW}=30$. Each panels shows different fidutial values of the GW amplitude $A_* = [10^{-11},\,10^{-9},\,10^{-7}]$.}
\end{figure*}

\begin{figure*}
\includegraphics[width=\columnwidth]{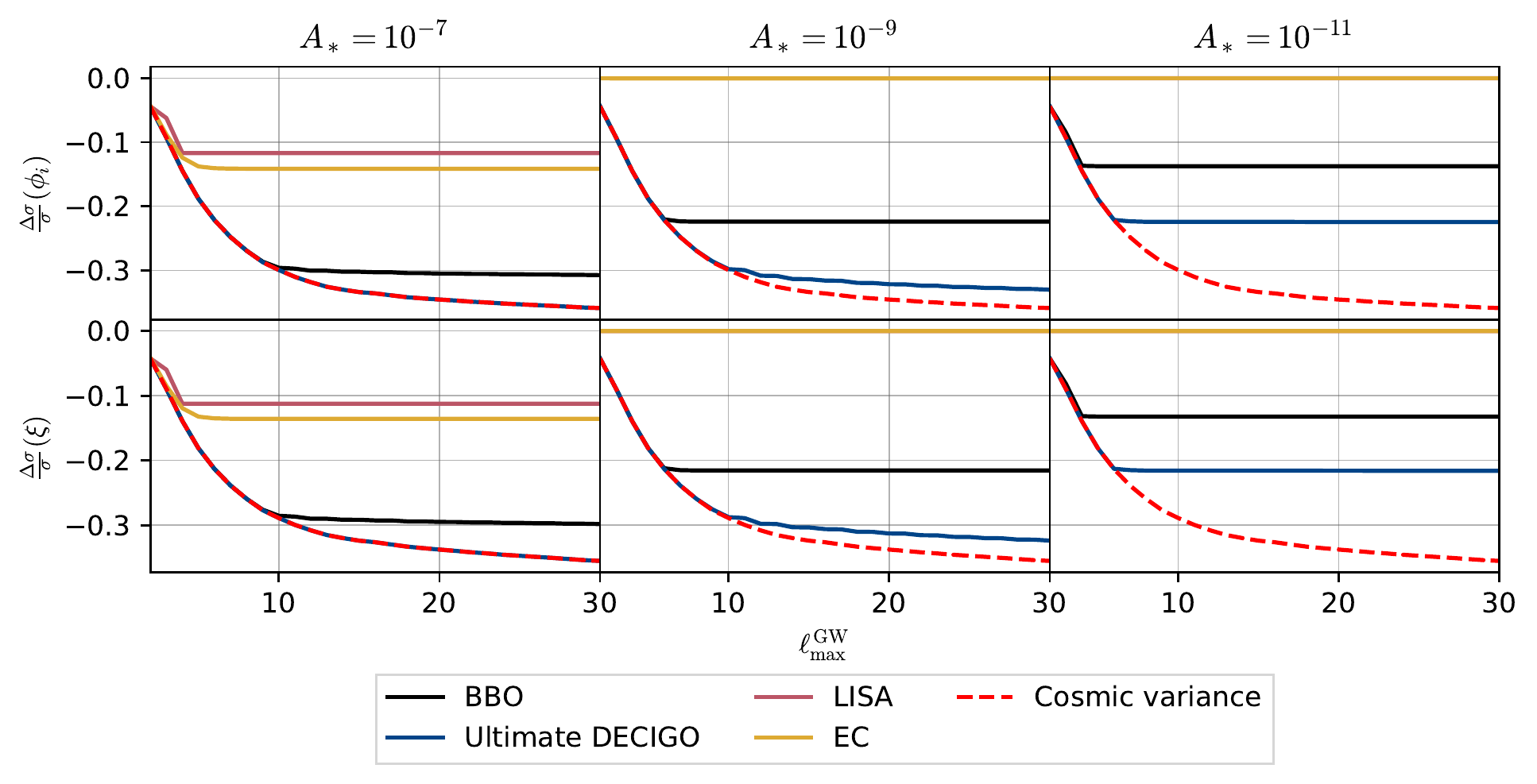}
\caption{\label{fig:sigmaNMC} Results for the NMC model. We plot the relative improvement $\Delta\sigma/\sigma\equiv (\sigma_{\rm CMB + SGWB}-\sigma_{\rm CMB})/\sigma_{\rm CMB}$ of the errors on $\phi_i$ and $\xi$.}
\end{figure*}

\subsection{Forecasts on the $\Delta N_{\rm eff}$ model}
We start by considering the $\Delta N_{\rm eff}$ model. We adopt the following fiducial cosmology \cite{Braglia:2021axy} : 
\begin{align}
&   100*\theta_s=1.0410,\,\,\,\, 100\, \omega_b=2.274,\, \,\,\,\,   \omega_c=0.1246,\notag\\
 &   \tau_{\rm reio}=0.058,\,\,\,\,\ln10^{10}A_s=3.063,\notag n_s=0.9786,\\&N_{\rm eff}=3.041.\label{eq:bestNeffR19}
\end{align}

The results of our Fisher analysis are shown in Fig.~\ref{fig:sigmaNeff}, where we plot the quantity $\Delta\sigma/\sigma\equiv (\sigma_{\rm CMB + SGWB}-\sigma_{\rm CMB})/\sigma_{\rm CMB}$, which measures the fractional improvement in the constraints that we gain by adding measurements of SGWB anisotropies to CMB ones. 
In addition to the results for different experiments, we also plot the relative error for the cosmic variance limited case (the red dashed lines). We can see that the error decreases when both $A_*$ and $\ell_{\rm max}^{\rm GW}$ increase. Indeed, by increasing the monopole amplitude $A_*$, the SNR becomes larger and, as we see in Fig.~\ref{fig:ClNeff}, the variations in the anisotropies spectra extend up to large multipoles, so larger values of $\ell_{\rm max}^{\rm GW}$ enhance the constraining power.

The only GWI that can reach the cosmic variance limit is Ultimate DECIGO in the case of a large amplitude of the monopole. Fig.~\ref{fig:sigmaNeff} shows that EC and LISA improve the constraints very little and only for a large monopole amplitude. The situation is more optimistic for BBO and DECIGO for which the improvement on the errors is a bit better and can be obtained even with smaller values of $A_*$. However, a realistic experiment would have maximum angular resolution of $\ell \sim15$~\cite{Contaldi:2020rht}, and in this case, the improvement in the error is never larger than $\sim10\%$.

\subsection{Forecasts on the NMC model}
For the NMC model we choose the following fiducial cosmology, see Table~III of Ref.~\cite{Abadi:2020hbr}:
\begin{align}
&  100*\theta_s=1.0420,\,\,\,\, 100\, \omega_b=2.247,\, \,\,\,\,   \omega_c=0.1192,\notag\\
 &   \tau_{\rm reio}=0.060,\,\,\,\,\ln10^{10}A_s=3.060,\notag n_s=0.9727,\\&\xi=-1/6,\,\,\,\, \phi_i  =0.297.\label{eq:paramsNMC}
\end{align}

Our results are shown in Fig.~\ref{fig:sigmaNMC}, where, as before, we show the relative improvement of the forecast error obtained by cross-correlating SGWB and CMB anisotropies with respect to the one by CMB anisotropies alone. The relative improvement in the constraints is very similar for both the coupling $\xi$ and the initial condition $\phi_i$. Unlike the $N_{\rm eff}$ model, the modifications induced by the NMC model are confined to very large scales (compare Figs.~\ref{fig:ClNeff} and \ref{fig:ClNE}), so the gain in considering larger multipoles is reduced compared to the one observed in the previous subsection. Indeed, as can be seen from the red dashed line, representing a CV limited GWIs, $\ell_{\rm max}^{\rm GW}=15$ is already enough to gain a $\sim30\%$ improvement in the errors wrt to the CMB ones.

In the case of LISA and EC, we need a large monopole amplitude of $A_*=10^{-7}$ to improve the constraints of $\sim10\%$, but no further improvements are seen for $\ell_{\rm max}^{\rm GW}\geq5$, because the noise $N_\ell$ becomes too large. On the other hand, for the same amplitude, both BBO and Ultimate DECIGO will decrease the errors by $\sim20\%$. For Ultimate DECIGO, the noise is so small that this also holds for a smaller amplitude of $A_*=10^{-9}$, while the noise of BBO is larger, allowing to map SGWB anisotropies only up to $\ell_{\rm max}^{\rm GW}=6$, and the relative improvement shrinks to $\sim 14\%$. For an even smaller amplitude $A_*=10^{-11}$, the SNR decreases, and BBO (Ultimate DECIGO) is able to map SGWB anisotropies only up to $\ell_{\rm max}^{\rm GW}=4\, (6)$, reducing the improvement to $\sim 10\%\, (14\%)$.

\subsection{Forecasts on the EDE model}
\begin{figure}
\includegraphics[width=\columnwidth]{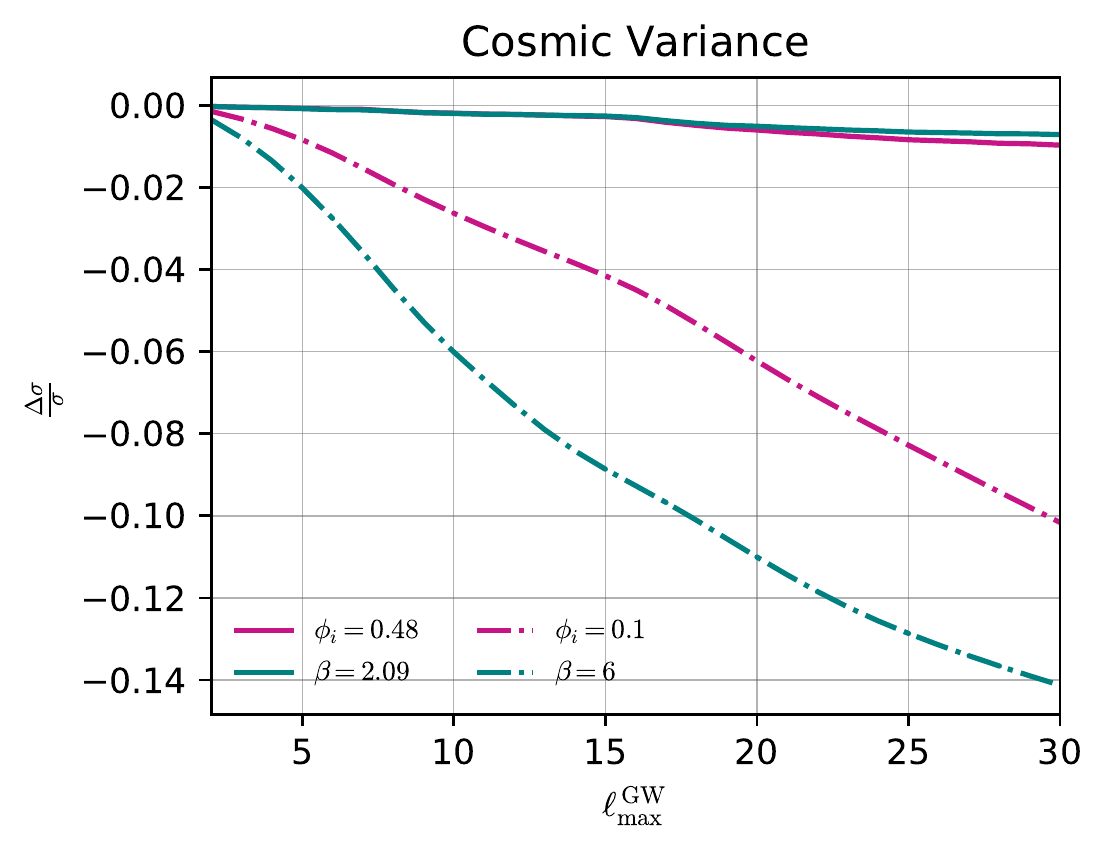}
\caption{\label{fig:sigmaRnR} Results for the EDE model. We plot the relative improvement $\Delta\sigma/\sigma\equiv (\sigma_{\rm CMB + SGWB}-\sigma_{\rm CMB})/\sigma_{\rm CMB}$ of the errors on $\phi_i$ and $\beta$. We only consider the illustrative case of a Cosmic Variance limited IGW and present the two fiducial sets of parameters discussed in the text.}
\end{figure}

For the EDE model we choose the following fiducial cosmology, see Table~II of Ref.~\cite{Braglia:2020auw}:
\begin{align}
&  100*\theta_s=1.0417,\,\,\,\, 100\, \omega_b=2.286,\, \,\,\,\,   \omega_c=0.1242,\notag\\
 &   \tau_{\rm reio}=0.059,\,\,\,\,\ln10^{10}A_s=3.059,\notag n_s=0.9813,\\&\phi_i=0.48,\,\,\,\,\beta=2.09.\label{eq:paramsEDE}
\end{align}
The former set of parameters represent an energy injection of about $6\%$ of the total energy density of the Universe located at the redshift $z_c=3390$ \cite{Braglia:2020auw}. As shown in Section~\ref{sec:EDE}, the SGWB anisotropies at $\ell\leq30$ depend very weakly on variations around this fiducial choice of parameters. Therefore, as expected, we find that the inclusion of SGWB anisotropies improves the error on the model parameter of less than $1\%$, as shown by the solid lines in Fig.~\ref{fig:sigmaRnR}, even if no noise is considered for the GWIs.

As an illustrative case, it is interesting to explore the case of earlier energy injection, which increases the imprints of EDE on the SGWB spectra as shown in Section~\ref{sec:EDE}. For fiducial values, we choose the $\Lambda$CDM best-fit parameters for the same choice of the dataset used for obtaining the values in Eq.~\eqref{eq:paramsEDE}, which are also given in Table~II of Ref.~\cite{Braglia:2020auw}:
\begin{align}
&  100*\theta_s=1.04229,\,\,\,\, 100\, \omega_b=2.265,\, \,\,\,\,   \omega_c=0.1178,\notag\\
 &   \tau_{\rm reio}=0.057,\,\,\,\,\ln10^{10}A_s=3.047, n_s=0.9719
\end{align}
and add a small EDE component corresponding to an energy injection of $\sim0.4\%$ at the redshift $z_c\sim1.6\times10^5$, obtained by setting $\phi_i=0.1$ and $\beta=6$.

In this case, as shown in Fig.~\ref{fig:ClNE}, since the energy injection occurs deep in the radiation era, the subsequent radiation-like redshifting of the averaged energy density of EDE contributes to the total expansion history, sizably affecting the SGWB anisotropies. For this reason, we can observe in Fig.~\ref{fig:sigmaRnR} that the error on the EDE parameters improves much more than in the previous case.

Although the scenario just discussed is already excluded by CMB data, it can be useful as an illustrative example of the capability of SGWB anisotropies to constrain the physics of the very early Universe. In particular, our findings suggest that they will be a valuable tool to constrain scenarios, such as those discussed in  Refs.~\cite{Caldwell:2018giq,DEramo:2019tit,Domenech:2020kqm}, where the equation of state of the Universe differ from the one of radiation at very large redshifts and for a limited amount of time, complementing the information encoded in the spectral shape of the monopole~\cite{Caldwell:2018giq,DEramo:2019tit}. We note that such scenarios do not leave any direct imprints on the CMB since the modified dynamics entirely occur well before the last scattering surface.

\section{Conclusions}
\label{sec:conclusions}
The anisotropies of the Stochastic Gravitational Wave Background (SGWB) have received increasing attention. The information they encode could be exploited to disentangle the contribution to the SGWB of astrophysical and cosmological processes, which are expected to contribute with distinct signature to the anisotropies. As a cosmological probe, the SGWB anisotropies induced by the propagation of GWs through the large-scale density perturbations \cite{Bartolo:2019oiq,Bartolo:2019yeu} have been recently shown to be sensitive to the imprints of the physics operating at very large redshift during the radiation dominated era \cite{DallArmi:2020dar}. It is therefore natural to maximize the information carried by SGWB anisotropies by cross-correlating it with the ones of the CMB, which are currently the best observable to constrain early Universe physics \cite{Adshead:2020bji,Malhotra:2020ket}. The advantage of considering GWs is that they decouple at the end of inflation, much earlier than the CMB photons, leading to a longer time intregration over the graviton's line of sight.

In this paper, we have explored the SGWB anisotropies produced by non-standard models of pre-recombination physics, and computed their cross-correlation with CMB anisotropies in temperature and polarization. To model the change in the early Universe cosmological history, we have considered three popular models that modify it by adding extra-relativistic degrees of freedom, as in \cite{DallArmi:2020dar}, a Non-Minimally coupled scalar (NMC) field, and an Early Dark Energy (EDE) component. Some of our main results are summarized in Figs.~\ref{fig:ClNeff}, \ref{fig:ClCC} and \ref{fig:ClNE} that show that the three models affect the theoretical spectra in different ways.

In order to quantify the capability of future planned Gravitational Wave Interferometers (GWIs) for constraining pre-recombination physics, we have performed a simple Fisher analysis including SGWB and CMB anisotropies and their cross-correlation. Since GWIs will only be sensitive to large angular scales due to their poor resolution, we have considered a large-scale CMB polarization experiment in our analysis and use the specifications of the LiteBIRD satellite. Figs.~\ref{fig:sigmaNeff} and \ref{fig:sigmaNMC} show the reduction of the error by the addition of GWIs interferometers to LiteBIRD. We have shown that the exact magnitude of this reduction depends on the model considered, and more importantly, on the GWI used and the monopole amplitude of the SGWB. As expected, the maximal improvement is obtained for experiments with better sensitivity and angular resolution.

Regarding the model dependencies, we have found that adding SGWB measurements to CMB ones improves the error on the parameters describing extra-relativistic degrees of freedom or early non-minimal couplings to gravity, while EDE models affect the anisotropies spectra only at large multipoles out of the reach of future GWIs. However, for illustrative purposes, we have also shown the case of a very early energy injection within EDE models, for which the improvement in the error on the model parameters is more optimistic. Our results suggest that SGWB are a useful tool to constrain variations of the equation of state deep in the radiation era \cite{Caldwell:2018giq,DEramo:2019tit,Domenech:2020kqm}.

Our analysis can be improved in several ways. First of all, in order to be able to cross-correlate SGWB and CMB spectra, we have considered a large-scale polarization CMB experiment. However, the models that we have analyzed could impact the high multipoles of the CMB spectra and therefore are better constrained by experiments that can map those scales, such as Planck or the future planned CMB S4. Such choice will be, however, complicated by the need to include CMB lensing spectra, which have a crucial impact on the acoustic structure of the CMB spectra. We leave this for future work. Another important point is that, for definiteness, we have estimated the angular spectra of SGWB anisotropies at a fixed reference frequency $f_*$ that we took, for each experiment, as the location of its sensitivity peak. As shown in Eq.~\eqref{eq:CellOmega}, though, the angular spectra are generically frequency dependent and follow the same dependence of the monopole. It would be interesting to exploit such typical behavior to distinguish anisotropies generated by the propagation of GWs in large scale density perturbations from other sources of anisotropies with a different dependence on the frequency and to explore better SGWB anisotropy estimators making use of the anisotropies at different frequencies.

\vspace{0.5cm}

{\bf Note added:} While this project was nearly complete, we became aware of Ref.~\cite{Ricciardone:2021kel}, where the cross-correlation between CMB temperature anisotropies and the cosmological and astrophysical SGWB was analyzed and used to produce constrained realization maps of the SGWB  anisotropies out of the CMB ones.

\vspace{0.5cm}

{\bf Acknowledgements}
MB would like to thank David Alonso for help with the schNell code and Dhiraj Hazra for discussions on Fisher forecasts. We thank the authors of Ref.~\cite{Ricciardone:2021kel} for sharing the draft of their paper with us. Numerical computations for this research were done on the Sciama High Performance Compute cluster, which is supported by the ICG, SEPNet, and the University of Portsmouth. The authors are supported by the Atracción de Talento contract no. 2019-T1/TIC-13177 granted by the Comunidad de Madrid in Spain. SK is partially supported by Japan Society for the Promotion of Science (JSPS) KAKENHI Grant no. 20H01899 and 20H05853.

\end{document}